\documentclass{article}
\usepackage{graphicx}
\usepackage{epstopdf} 
\usepackage{bm}
\usepackage{amsmath}
\usepackage{cite}
\usepackage{authblk}

\begin{document}

\oddsidemargin -5mm \evensidemargin -5mm

\title{\vspace{-10ex} Single molecule imaging with longer x-ray laser pulses }

\author[1]{Andrew V. Martin}
\author[1]{Justine K. Corso}
\author[2,3]{Carl Caleman}
\author[2]{Nicusor Timneanu}
\author[1]{Harry M. Quiney}
\affil[1]{ARC Centre of Excellence for Advanced Molecular Imaging, School of Physics, University of Melbourne, Parkville, Victoria 3010, Australia}
\affil[2]{Department of Physics and Astronomy, Uppsala University, Box 516, SE-751 20 Uppsala,
Sweden}
\affil[3]{Center for Free-Electron Laser Science, DESY, Notkestrasse 85, DE-22607 Hamburg, Germany}

\maketitle


\begin{abstract}
During the last five years, serial femtosecond crystallography using x-ray laser pulses has developed into a powerful technique for determining the atomic structures of protein molecules from micrometer and sub-micrometer sized crystals. One of the key reasons for this success is the ``self-gating" pulse effect, whereby the x-ray laser pulses do not need to outrun all radiation damage processes. Instead, x-ray induced damage terminates the Bragg diffraction prior to the pulse completing its passage through the sample, as if the Bragg diffraction was generated by a shorter pulse of equal intensity. As a result, serial femtosecond crystallography does not need to be performed with pulses as short as 5--10 fs, as once thought, but can succeed for pulses 50--100 fs in duration. We show here that a similar gating effect applies to single molecule diffraction with respect to spatially uncorrelated damage processes like ionization and ion diffusion. The effect is clearly seen in calculations of the diffraction contrast, by calculating the diffraction of average structure separately to the diffraction from statistical fluctuations of the structure due to damage (``damage noise"). Our results suggest that sub-nanometer single molecule imaging with 30--50 fs pulses, like those produced at currently operating facilities, should not yet be ruled out. The theory we present opens up new experimental avenues to measure the impact of damage on single particle diffraction, which is needed to test damage models and to identify optimal imaging conditions. 
\end{abstract}

\section{Introduction}

X-ray free-electron laser (XFEL) pulses are envisioned to probe the structures of radiation-sensitive samples, like biological molecules, by outrunning radiation damage processes\cite{Neutze2000}. Current facilities, however, produce their brightest pulses with durations of the order of tens of femtoseconds\cite{LCLS,Ishikawa2012}, which is sufficient time for ionization to become widespread and for ions to move several {\AA}ngstr{\"o}ms\cite{Caleman2009,Caleman2011}. In spite of this, the first applications of XFELs to serial crystallography have been highly successful\cite{Chapman2011,Boutet2012}. It turns out that even for longer pulses ($\sim$ 50--100 fs), Bragg diffraction probes the undamaged structure in the first few femtoseconds of the pulse-sample interaction, turning off at later times when radiation damage distributes the diffraction signal as a diffuse background\cite{Barty2012}. In this way, XFEL Bragg diffraction is effectively gated by damage because expected number of photons scattered to a Bragg peak is equivalent to that produced by a shorter pulse with the same intensity. 

Despite the great progress in coherent imaging using XFEL sources, the holy grail - atomic resolution of a single (non-crystalline) biomolecule \cite{Neutze2000} - has not yet been realized. Nevertheless, the potential reward for success has kept this pursuit at the forefront of research in XFEL imaging science. One of the limiting factors is radiation damage. For non-crystalline samples, diffraction from the undamaged structure is not enhanced by periodicity and is mixed indistinguishably with the diffraction of a damaged structure. This is seemingly a major setback for the prospects of developing 3D single particle imaging into a high resolution technique for single molecules. For example, Hau-Riege et al.\cite{HauRiege2005} found that radiation damage causes large discrepancies with the ideal diffracted intensities, which led them to conclude that pulses must be no more than a few femtoseconds long to avoid severe resolution loss. A more recent study with more detailed scattering models reached a similar conclusion\cite{Ziaja2012}. However, these studies assessed feasibility with metrics inspired by crystallography whose suitability for single molecule imaging is disputed\cite{Quiney2011}. Without accounting in detail for the way that structural information is extracted from single molecule diffraction data, the issue of damage limits for single molecule imaging remains inconclusive.

One of the most actively pursued routes to single molecule imaging involves measuring thousands of copies of a molecule one by one. The resulting data is extremely noisy and the molecular orientations are not known. The issue of molecular orientation must be resolved to assemble a 3D dataset, which can be performed by several algorithms \cite{LohEMC2009,Fung2009,Giannakis2012,Kassemeyer2013}. The hallmark of these methods is that they are able to cope with signals as low as 0.01 photons per Shannon-Nyquist pixel \cite{Tegze2012}. After the 3D dataset has been assembled, the atomic structure is recovered via coherent diffractive imaging methods\cite{Marchesini2007}. 

\begin{figure}
\begin{center}
\includegraphics{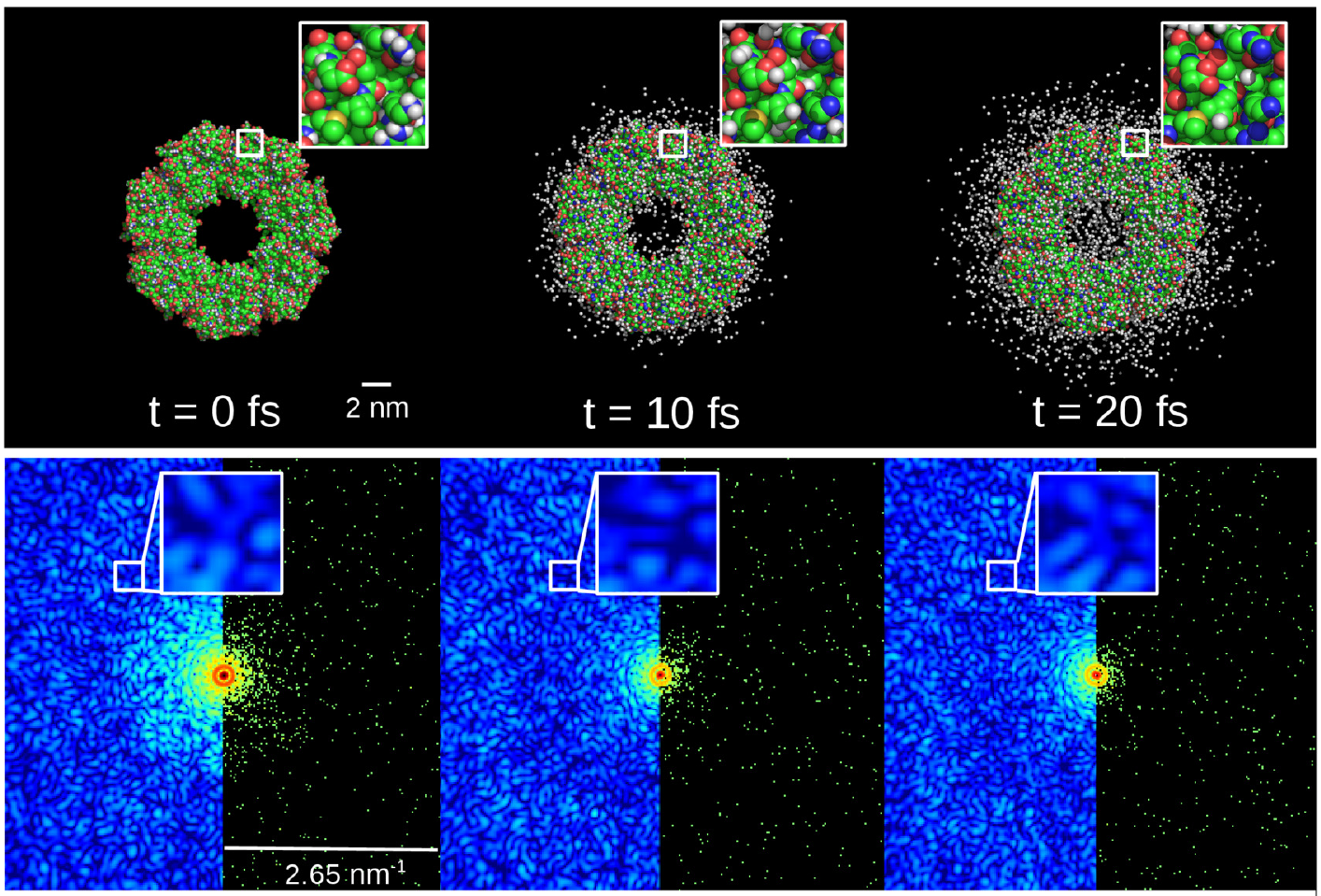}
\end{center}
\caption{ A graphical representation of ion diffusion in GroEL, where ion locations are chosen stochastically using the time-dependent temperature. Simulation parameters are: 8 keV; 5.0 $\times$ 10$^{20}$ W cm$^{-2}$; and 100 nm pulse diameter. Ionized hydrogen (white) moves much faster than ions of other elements. The diffraction pattern for each time point is shown below and was generated by randomly assigning each atom an ionization state and a displacement according to a rate-equations model described in Appendix \ref{app:model}. Large changes to the speckle structure are predicted at high resolution, as shown by the enlarged inset regions. The effect of shot-noise is shown on the right half of each diffraction image. }
\label{fig:diffraction_and_explosion}
\end{figure}

The crucial information needed to resolve the unknown orientations, and finally the structure, is contained in the modulations of diffraction signal arising from interference between different atoms, often called ``speckles" (see Fig. \ref{fig:diffraction_and_explosion}). Radiation damage changes the structure of the sample dynamically such that the final diffraction pattern is the sum of the diffraction from many modified structures, each with a different distribution of ions and ion displacements. It has been shown that averaging the diffraction over different molecular configurations\cite{Maia2009} lowers the speckle contrast relative to the mean scattering intensity within each resolution shell. We expect radiation damage to cause a similar loss of contrast. Not only is the amplitude of the speckle structure reduced, but speckle structure also fluctuates from shot-to-shot due to damage, in addition to the fluctuations due to changing orientation and shot-noise. We will use the term ``damage noise" to refer to these fluctuations of speckle structure due to damage. So far damage noise has not been considered in studies of 3D dataset assembly. Here we present calculations of damage noise per diffraction pattern due to spatially uncorrelated damage processes, which include ionization and ion diffusion but not the Coulomb explosion of the molecule. An analysis of damage noise as a function of pulse duration reveals a gating effect in single molecule diffraction, whereby long pulses measure an equivalent amount of information about the average structure to shorter pulses of the same intensity. Theoretical predictions of damage noise are also the first step to understanding how orientation determination and 3D data assembly can be performed with data affected by radiation damage. 

An alternative to alignment via post-processing is to experimentally align isolated gas-phase molecules, e.g. via quantum-state-selection methods \cite{Kupper2014,Stern2014}. A great advantage of this approach is that multiple molecules can be illuminated simultaneously, increasing signal-to-noise and, as supported by the work here, reducing the impact of damage. These methods have been demonstrated only for small (2,5-diiodo-benzonitrile) molecules so far \cite{Kupper2014,Stern2014} and extensions to larger molecules are being actively pursued. If the molecules are aligned experimentally, the self-gating effect still applies.  Radiation damage modifies each molecule in the beam uniquely and stochastically, so that multiple damage scenarios are averaged in a single diffraction measurement in an analogous way to crystallography. This increases the signal with respect to damage noise as well as shot noise. The self-gating effect ensures that such benefits from using multiple aligned molecules are not lost entirely by using x-ray pulses longer than 10 fs.

Once the 3D data assembly has been performed, damage will still have a residual effect on the resulting 3D diffraction volume. Damage reduces the contrast in the averaged diffraction volume\cite{Quiney2011}, and depending on the theoretical perspective, also contributes a background\cite{Lorenz2012}. Promisingly the reduction in contrast can be accounted for during structure determination by treating the sample in terms of a small number of structural modes\cite{Quiney2011}. The background contribution is expected to be small for hard X-rays at beam conditions currently available.  

In addition to analysing the damage noise, we show how the mean and standard deviation of the diffraction signal can be combined into a sensitive measure of damage. An advantage of the measure we propose is its sensitivity to both ionization and ion motion, whereas the mean signal alone depends only on ionization. There is a  need to measure damage experimentally and provide some validation and clarification for theoretical damage modelling. Many different types of damage models have been developed, based on rate-equations\cite{HauRiege2004}, molecular dynamics \cite{Neutze2000, Jurek2004b} or plasma theory\cite{Caleman2009}, and each has specific advantages and disadvantages. For example, molecular dynamics models can keep track of specific ion trajectories, but are only computationally tractable for small molecules \cite{Neutze2000}. Rate equations models can simulate damage large molecules, but ignore information about ion motion on atomic length scales \cite{HauRiege2004}. Experimental measurements of damage will provide valuable feedback on our theoretical understanding of the interaction between XFEL pulses and biomolecules, which is needed to develop single molecule imaging techniques.

\section{The effect of radiation damage on diffraction contrast}

The goal of single molecule imaging is to recover the initial position \textbf{R} of each atom in the sample. For simplicity, we will give equations for the case of a single atomic species, noting that the generalization to multiple atomic species is similar to that found in Ref. \cite{Quiney2011}. The intensity of a single measurement of a single molecule can be written
\begin{equation}
I(\textbf{q}) = r_e^2 P(\textbf{q}) d\Omega I_0 \left[ \sum_{i=1}^N A_{i}(q) + 2 \sum_{i=1}^N \sum_{j=1}^{i-1} B_{ij}(\textbf{q}) \right] \;,
\label{eq:I}
\end{equation}
where $\textbf{q}$ is the scattering vector with magnitude $q$, $d\Omega$ is the solid-angle term, $r_e$ is the classical electron radius, $N$ is the number of atoms and $P(\textbf{q})$ is a polarization term that will be ignored in this discussion. To simplify mathematical notation, we assume the incident intensity takes a uniform value $I_0$ for the duration of the pulse. We have defined
\begin{equation}
A_{i}(q) = \int_0^T |f_i(q,t)|^2  dt \;
\label{eq:Ai}
\end{equation}
and 
\begin{align}
B_{ij}(\textbf{q}) &=  \int_0^T f_i(q,t) f_j(q,t) \cos[ 2 \pi \textbf{q} \cdot (\textbf{R}_i - \textbf{R}_j + \bm{\epsilon}_i(t) - \bm{\epsilon}_j(t))] dt \;,
\label{eq:Bij}
\end{align}
where $\bm{\epsilon}_i(t)$ is the displacement of the $i^{th}$ atom from its initial position and $T$ is the duration of the pulse. For a single two-dimensional measurement, it is understood that  $\textbf{q}$ is sampled at points on the Ewald sphere, but in general we will use $\textbf{q}$ to be a general three-dimensional vector and $I(\textbf{q})$ is a three-dimensional function.  The atomic scattering factor $f(q,t)$ depends upon the ionization state of the atom, which changes as a function of time. The ionic scattering factors can be calculated using Slater orbitals\cite{Slater1930} and we use $f_0(q)$ to denote the atomic scattering factor of the unionized atom. We assume that the probability of an ion having a particular ionization state at time $t$ is independent of where that atom is located in the sample. Although the ionization state as a function of time is different for each atom, statistically atoms of the same atomic species are assumed to be equivalent. We write $A(q)$ and $B(q)$ as a function of the magnitude of the scattering vector, $q$, because we assume the atomic scattering factors are spherically symmetric. 

Consider an ensemble of 2D diffraction measurements, each with a unique damage scenario. For 3D imaging, the data needs to be assembled into a 3D intensity volume using an algorithm that accounts for the unknown molecular orientations. The desired solution of the algorithm is an average intensity, where each 2D measurement is correctly placed according to orientation and the different damage scenarios are averaged. As shown in Appendix \ref{app1}, the average intensity can be written in the form 
\begin{equation}
\langle I(\textbf{q}) \rangle = r_e^2 P(\textbf{q}) d\Omega I_0 \left[ N A(q) + 2 B(q) \sum_{i=1}^N \sum_{j=1}^{i-1} \cos[ 2 \pi \textbf{q} \cdot (\textbf{R}_i - \textbf{R}_j)] \right] \;,
\label{eq:Iav}
\end{equation}
where we have
\begin{equation}
\langle A_{i}(q) \rangle = A(q) \equiv I_0 \int_0^T \langle |f(q,t)|^2 \rangle  dt 
\label{eq:Aav}
\end{equation}
and 
\begin{equation}
\langle B_{ij}(\textbf{q}) \rangle = B(q) \cos[ 2 \pi \textbf{q} \cdot (\textbf{R}_i - \textbf{R}_j)] \;,
\label{eq:Bijav}
\end{equation}
where
\begin{equation}
B(q) \equiv \int_0^T  \langle f(q,t) \rangle^2 e^{-4\pi^2 q^2 \overline{\bm{\epsilon}}(t)^2} dt
\end{equation}
and $\overline{\bm{\epsilon}}(t)$ is the root mean square (rms) displacement of an ion as a function of time. 

If the analysis is restricted to damage processes that are random and spatially uncorrelated, then we can treat the terms $A_{i}(q)$ and $B_{ij}(\textbf{q})$ as random variables and study the effect of damage statistically. We also treat the initial atomic positions $\textbf{R}_i$ as random with a uniform probability distribution, as is done in crystallography to analyse the statistics of Bragg intensities (Wilson statistics) at high scattering angles ($q > 0.33 $nm$^{-1}$) \cite{Huldt2003}. Both ionization and ion diffusion can be treated within this framework and, as we will show, are both involved in a self-gating pulse effect. Expansion of the molecule by Coulomb forces is not covered by the statistical treatment presented here, but is discussed further below.

The second term on the right-hand side of Eq. \eqref{eq:Iav} is sensitive to the atomic positions and accounts for the contrast in the average diffraction pattern. We can treat this information as the ``signal" we aim to measure. The contribution each atom makes to the signal is proportional to $B(q)$, which is equal to the standard deviation of the diffraction in the merged 3D dataset divided by the number of atoms. The mean shot noise level, denoted by $\sigma_N$, is proportional to the square root of the intensity. We can estimate the mean shot noise level by considering the mean diffracted intensity in a shell of constant $q$, which can be derived by integrating Eq. \eqref{eq:Iav} and is proportional to $A(q)$. When the signal is compared to the noise, the proportionality constants have no influence on the interpretation, so we drop them for simplicity and write  
\begin{equation}
\sigma^2_N(q) =  A(q)  \;.
\label{eq:sigN}
\end{equation}
In addition to shot noise, there is the damage noise due to the variations in how the damage manifests in each measurement. One contribution to the damage noise is the fluctuation of $A_{i}(q)$, which is characterized by the standard deviation of $A_{i}(q)$, which we denote by $\sigma_A(q)$. The second contribution to damage noise is the deviation of $B_{ij}(\textbf{q})$ from the average speckle $B(\textbf{q})$, which has a standard deviation $\sigma_B(q)$. The term $\sigma_B(q)$ is given by the difference between the standard deviation of the second term on the right-hand side of Eq \eqref{eq:I} minus the standard deviation of the second term on the right-hand side of Eq \eqref{eq:Iav}. In Appendices \ref{sec:app:sigA} and \ref{sec:app:sigB}, we provide derivations of $\sigma_A(q)$ and $\sigma_B(q)$ that give the following results:
\begin{align}
\sigma^2_A(q) = \int^T_0 \int^T_0 \left[ \langle f^2(q,t) f^2(q,t') \rangle - \langle f^2(q,t)\rangle \langle f^2(q,t') \rangle \right] dt dt'
\label{eq:sigA}
\end{align}
and
\begin{align}
\sigma^2_B(q) = \int^T_0 \int^T_0 \Big[  &\langle f(q,t) f(q,t') \rangle^2 e^{-4\pi^2 q^2 |\overline{\bm{\epsilon}}^2(t, t')|} 
\nonumber \\
&\qquad	- \langle f(q,t) \rangle^2 \langle f(q,t') \rangle^2 e^{-4\pi^2 q^2 \overline{\bm{\epsilon}}(t)^2} e^{-4\pi^2 q^2 \overline{\bm{\epsilon}}(t')^2} \Big] dt dt' \;.
\label{eq:sigB}
\end{align}
By comparing the size of the signal to the size of the shot-noise and damage-noise levels, we can gauge how much information is contained by each measurement about the molecule's structure. Here we will study how the diffraction pattern varies as a function of pulse duration and pulse energy. We propose the following signal-to-noise ratio to characterize the diffraction: 
\begin{align}
SNR_{ND}(q)  = \frac{N B(q)}{\sqrt{N \sigma^2_A(q) + N^2 \sigma^2_B(q) + N \sigma^2_N(q)}} \;.
\label{eq:SNR}
\end{align}
It is also interesting to compare the signal to the damage noise directly, ignoring shot-noise, with the following ratio:
\begin{align}
SNR_D(q)  = \frac{N B(q)}{\sqrt{N\sigma^2_A(q) + N^2\sigma^2_B(q) }} \;.
\label{eq:SNRD}
\end{align}

To estimate $SNR_{ND}(q)$ and $SNR_D(q)$, we need to calculate statistical averages of the scattering factor, e.g. $\langle f(q,t) \rangle$, $\langle f^2(q,t)\rangle$ etc, which in turn depend on the expected number of ions in each ionization state as a function of time. To calculate $B(q)$ and $\sigma_B(q)$ we also need to know the ion temperature as a function of time. These parameters can be calculated by many of the damage models reported in the literature so far, like molecular dynamics models \cite{Neutze2000, Jurek2004b} and hydrodynamic (rate-equations) models\cite{HauRiege2004,Scott2001, Caleman2011}. Here we will present the results of a rate equations model to investigate single-molecule diffraction contrast and to explore the extent to which there is a self-gating pulse effect in single molecule diffraction.

The term that has not been calculated before is the correlation between the scattering factor at different time points, e.g. $\langle f(q,t) f(q,t')\rangle$, which is needed to calculate the damage noise levels. To calculate these correlations we need to know the condition probability $P( f_n(q,t') | f_m(q,t) )$, which gives the probability of an ion being found in ionization state $n$ at time $t'$ given that it was in ionization state $m$ at time $t$. We have developed a way of calculating these conditional probabilities, and hence the damage noise. First the damage simulation is carried out generating the populations of ion states at all time points and the transition rates between ion states are stored as a function of time. Starting with the mean ion population of state $m$ at time $t$, the stored transition rates can be used to generate the fraction of these atoms in ionization state $n$ at all later time points $t' > t$, from which the conditional probabilities can be readily inferred.

We use a damage model based on a rate-equations model\cite{HauRiege2004}, which is extended to include ion diffusion using the methods from a non-local thermal equilibrium plasma model\cite{Scott2001,Caleman2011}. The details of the model are given in Appendix \ref{app:model}. As we closely follow the methods of Refs. \cite{HauRiege2004,Caleman2011}, we expect the results and the validity our model to be similar. As we will show, there are sufficient physical processes in our model to illustrate the self-gating pulse effect in single molecule diffraction. 

All statistical quantities are given as weighted averages over the light elements (H,C,N,O). Sulphur was included in the rate-equations model of damage, but was excluded from the average of statistical diffraction quantities, like $A(q)$, $B(q)$ and $\sigma_B(q)$, because it is computationally intensive. Sulphur has a much larger number of possible electron configurations, and averages that depend on two time variables [e.g. $\sigma_B(q)$] took too long to compute for the range of beam conditions we study here. Since there are of the order of 100 sulphur atoms and 10$^4$ light atoms, our main conclusions are not expected to be affected by neglecting the diffraction from sulphur.

We have set up our simulations using the chemical composition and size of GroEL. This chaperonin molecule is a candidate for first tests of single molecule imaging because it survives intact in mass spectrometry experiments\cite{Rostrom1999}, which subject the molecule to similar conditions to injection at XFEL. It is also of sufficient size to scatter around $10^4$ photons per diffraction pattern, as shown in Fig. \ref{fig:diffraction_and_explosion}. 

\begin{figure}
\begin{center}
\includegraphics{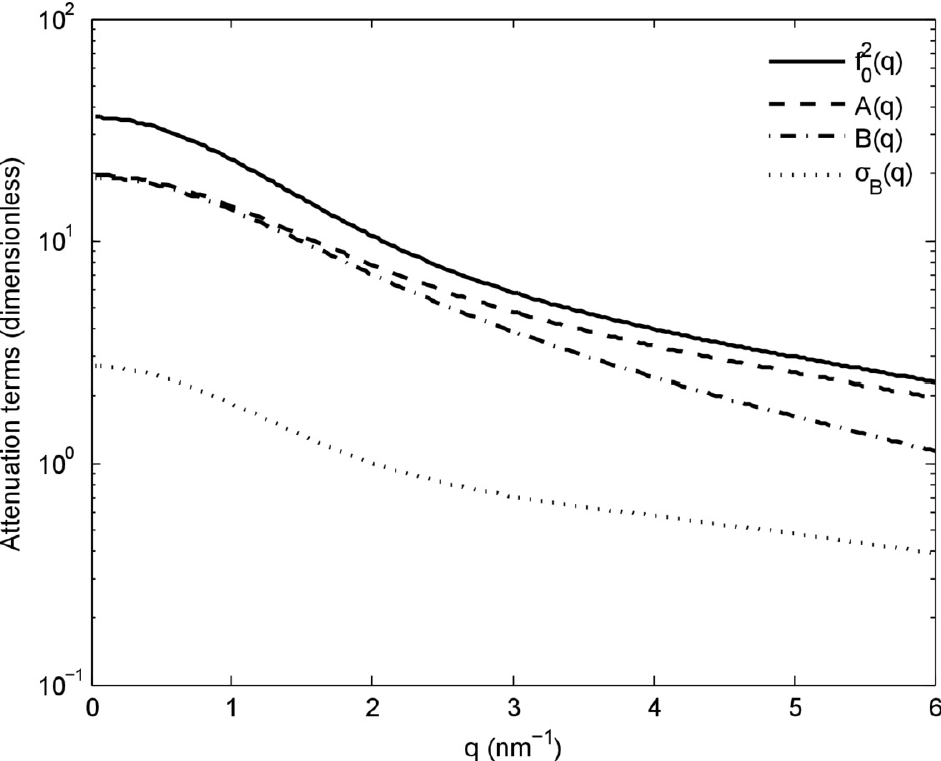}
\end{center}
\caption{ The effects of damage on the atomic structure factor. The term $f_0(q)$ is the undamaged atomic scattering factor for an unionized carbon atom, $A(q)$ is proportional to the mean intensity per carbon atom at each resolution shell, $B(q)$ is proportional to the speckle contrast for carbon and $\sigma_B(q)$ is the standard deviation of the shot-to-shot fluctuations of the speckle due to damage. When there is no damage $A(q)$ and $B(q)$ are equal to $f^2_0(q)$. The simulation parameters were 8 keV photon energy, 40 fs pulse duration, 2 mJ pulse energy and spot size of 100 $\times$ 100 nm$^2$.}
\label{fig:selfgating_AB}
\end{figure}

Simulations were performed at 8 keV photon energy ($\sim$0.155 nm wavelength) which is sufficient resolution for structural biology and similar to that demonstrated in simulation studies of single molecule imaging\cite{Tegze2012}. The principal effects of damage on molecular diffraction can be seen in Fig. \ref{fig:selfgating_AB}, which shows a simulation for a pulse duration of 40 fs, beam intensity of $5\times10^{20}$  W cm$^{-2}$ (corresponding to a 2 mJ pulse) and a 100 $\times$ 100 nm$^2$ spot size. Without damage $A(q)$ would be equal to $f_0^2(q)$, but with damage it is reduced, attenuating the mean intensity by the same amount. The attenuation occurs at all resolutions, but is a greater fraction of the original signal at lower resolutions. The term $B(q)$ is lower than $A(q)$ because of the effects of ion motion, and the discrepancy is more pronounced at higher resolution. The deviations between $A(q)$ and $B(q)$ are important for accurate structure retrieval methods\cite{Quiney2011}. In this case, the most significant damage noise term $\sigma_B(q)$ is lower than $B(q)$ across all resolutions, indicating that even for pulse durations as long as 40 fs damage noise does not exceed the signal from the average molecular structure.

\begin{figure}
\begin{center}
\includegraphics{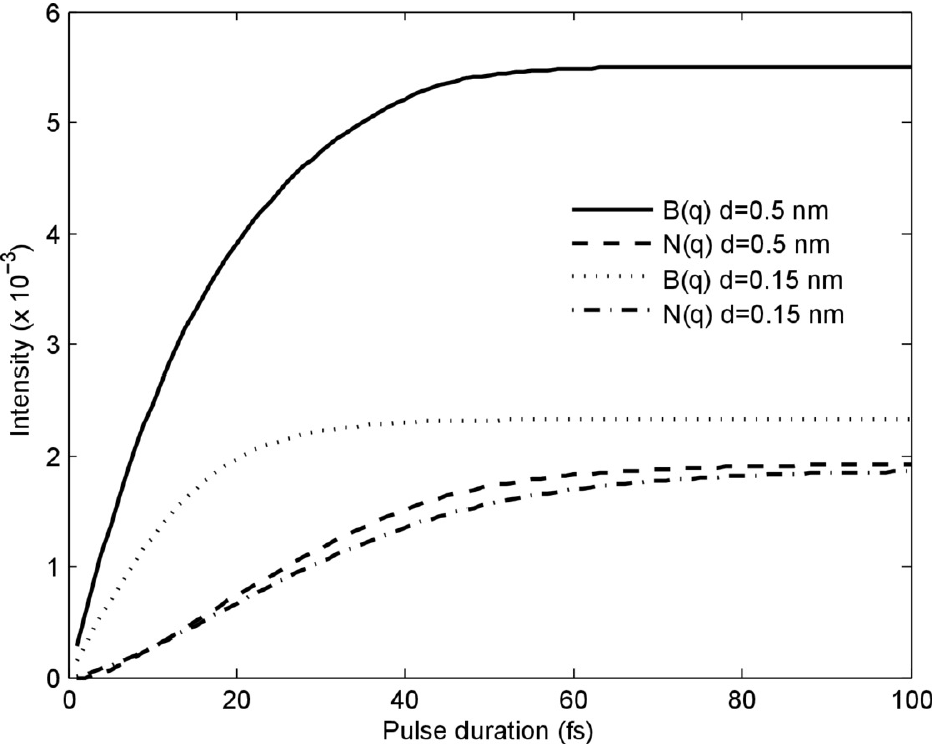}
\includegraphics{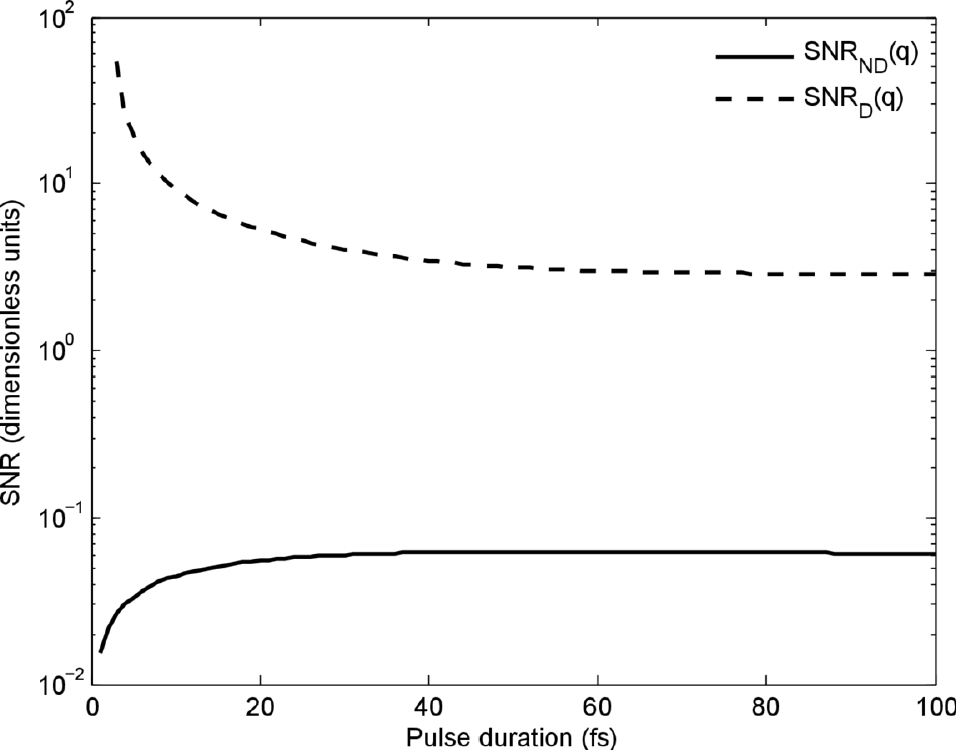}
\end{center}
\caption{ (a) Scattering and noise levels (due to damage only) as a function of pulse duration for constant incident intensity ($5 \times 10^{20}$ W cm$^{-2}$) at 8 keV photon energy and 100 $\times$ 100 nm$^2$ spot size. $B(q)$ is proportional to the speckle contrast and we define $N(q) \equiv \sqrt{\sigma^2_A(q)/N + \sigma^2_B(q)}$, which is the denominator in Eq. \eqref{eq:SNRD} and measures the average contribution to the damage noise per atom.  (b) Signal-to-noise ratios with and without shot noise for a resolution of 0.15 nm.  }
\label{fig:selfgating_SNR}
\end{figure}

To illustrate the self-gating pulse effect in single molecule diffraction, we plot $B(q)$ as a function of pulse duration for a constant photon energy (8 keV) and constant beam intensity ($5\times10^{20}$  W cm$^{-2}$). We see in Fig. \ref{fig:selfgating_SNR}(a) that the signal level at 0.15 nm resolution steadily rises until it plateaus at a maximum value at around 20 fs. The signal at lower resolution accumulates for longer pulse times. Interestingly the noise due to radiation damage also rises non-linearly, accumulating at slower rate at longer pulse times. This is because the random distribution of ions in the sample has smaller variation when the bound electrons are almost entirely depleted from each ion. The signal-to-noise ratios, shown in  Fig. \ref{fig:selfgating_SNR}(b), show strikingly that shot-noise has a much greater effect than damage noise. Although $SNR_D(q)$ improves greatly for short pulses ($<$5 fs), $SNR_{D+N}(q)$ maximizes when the signal $B(q)$ maximizes at around 20 fs. 

\begin{figure}
\begin{center}
\includegraphics{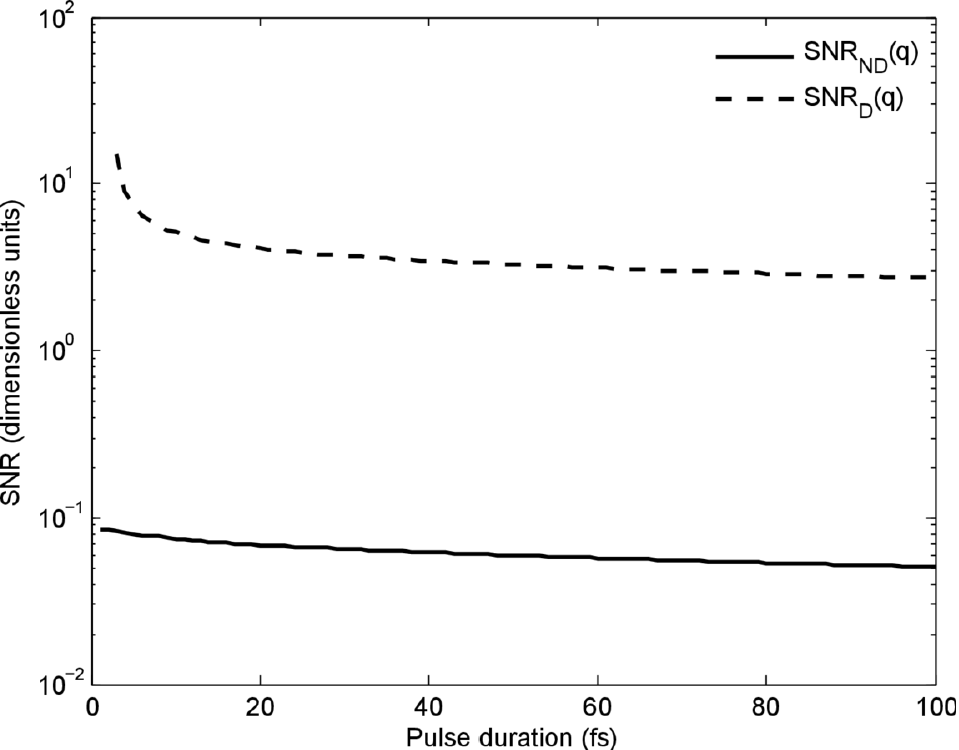}
\end{center}
\caption{ Maximum signal-to-noise ratios with and without shot noise for a resolution of 0.15 nm for 8 keV photon energy, 100 $\times$ 100 nm$^2$ spot size and constant pulse energy of 2 mJ. }
\label{fig:pulselength_SNR}
\end{figure}

The results are interesting when there is trade-off experimentally between pulse duration and pulse energy. For example, the LCLS can produce 2 mJ pulses with pulse durations of 30--50 fs for hard x-rays \cite{LCLS}. Sub 5 fs pulses can be produced by the LCLS using a low charge method or a slotted foil method, but at the expense of around a factor of ten in pulse energy. Given such a choice, the analysis presented here suggests that the gain in signal from a longer pulse with higher pulse energy compensates for the increase in damage. We note though, this conclusion only applies to spatially uncorrelated damage processes like ionization and ion diffusion (not a Coulomb explosion). Figure \ref{fig:pulselength_SNR} shows that $SNR_{D+N}(q)$ and $SNR_D(q)$ have a weak dependence on pulse duration at constant pulse energy. This suggests that maximizing pulse energy has a greater influence on the success of single molecule imaging than pulse duration with respect to the spatially uncorrelated damage mechanisms considered here. 

If multiple molecules were simultaneously aligned and exposed to the x-ray pulse (as described in the Introduction), we would still expect a gating effect qualitatively similar to that shown in Fig. \ref{fig:selfgating_AB}. However, we would expect $SNR_{D+N}(q)$ and $SNR_D(q)$ to scale as $\sqrt{N_{\rm mol}}$, where $N_{\rm mol}$ is the average number of molecules in the beam for each exposure. This is because the signal is proportional to $N_{\rm mol}$, while standard deviations of the damage noise and shot noise scale as $\sqrt{N_{\rm mol}}$. This analysis is missing the additional fluctuations due to the coherent interference between molecules, which have been considered in the context of angular correlation methods \cite{Kirian2012}.

\section{A method of measuring damage experimentally}

\begin{figure}
\begin{center}
\includegraphics{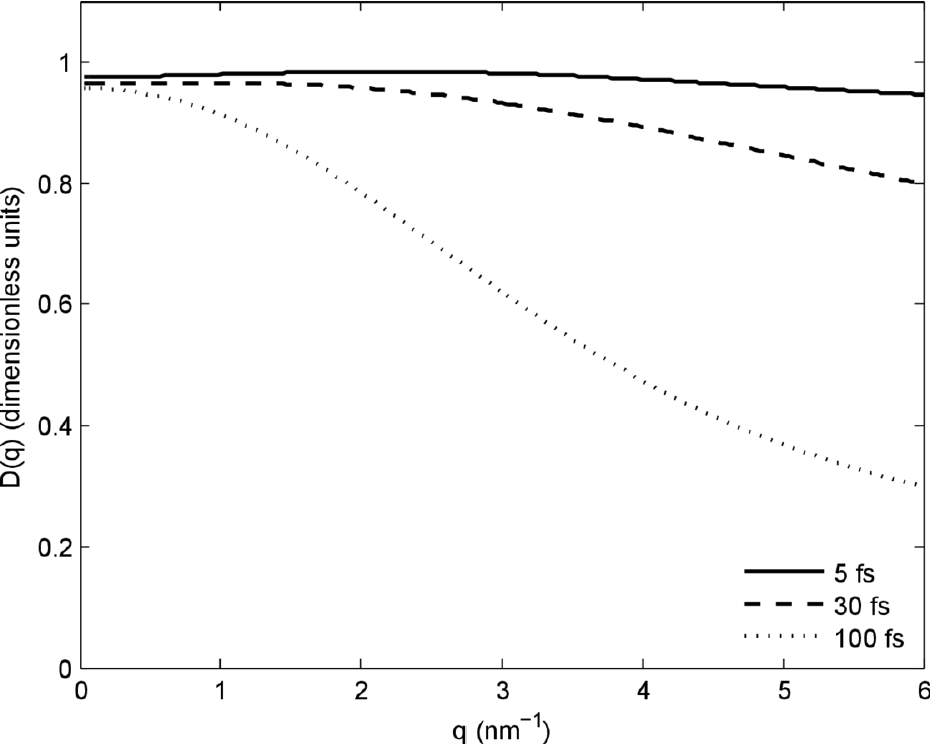}
\end{center}
\caption{ The function $D(q)$ for different pulse durations for 8 keV photon energy, 100 $\times$ 100 nm$^2$ spot size and constant pulse energy of 2 mJ.  }
\label{fig:pulselength_g2}
\end{figure}

The statistical analysis of diffraction contrast can be used to measure the amount of damage in single molecule experiments. The average change to the atomic structure factors, characterized by $A(q)$, can be readily measured by summing diffraction patterns. This provides some information about ionization levels but not ion motion. There is more information to be gained by analyzing the fluctuations of the diffraction signal. It is not convenient to measure $SNR_{D+N}(q)$, because $B(q)$ cannot be measured directly without resolving the issue of unknown orientations and assembling a 3D dataset, effectively accomplishing a full imaging experiment. An experimentally simpler proposition, which is independent of the imaging experiment, is to measure the standard deviation of the signal within each resolution ring, averaged over all of the measured diffraction patterns. The standard deviation is proportional to $\langle B_{ij}^2(q) \rangle$ and is a measure of the speckle contrast. It will contain both contributions from the average structure of the sample and the damage noise. Unfortunately it is not clear how to separate those two contributions experimentally. Nevertheless, the standard deviation is a sensitive measure of any dynamical change in the sample structure because it will drop relative to the mean scattering signal, as has been shown for averages of molecular conformation \cite{Maia2009}. To isolate the effect of damage-induced structural change, we create a measure that first subtracts the expected contribution of shot noise, which is equal to $ \mu_\textrm{pix}(q)$, and then normalizes by the mean intensity as follows:
\begin{equation}
D(q) = \frac{\sigma_{\textrm{pix}}^2(q) - \mu_\textrm{pix}(q)}{\mu^2_\textrm{pix}(q)} \;,
\end{equation}
where $\mu_\textrm{pix}(q)$ is the average intensity at a pixel in resolution ring $q$ averaged over the whole dataset and $\sigma_{\textrm{pix}}(q)$ is the corresponding standard deviation. The mean and standard deviation are calculated from the ensemble of experimental data of molecules measured individually in random orientations. It possible to show that
\begin{equation}
D(q) \approx \frac{\langle B_{ij}^2(q) \rangle}{ A^2(q)} \;,
\end{equation}
where $\langle B_{ij}^2(q) \rangle$ is given in Appendix \ref{sec:app:sigB}. It is possible to show  that $0 < D(q) < 1$, because $\langle f(q,t) f(q,t') \rangle^2 < \langle f^2(q,t) \rangle \langle f^2(q,t') \rangle$. Figure \ref{fig:pulselength_g2} shows $D(q)$ for variations of pulse duration at constant pulse energy (2 mJ). The large variations at high scattering angle indicate the sensitivity of $D(q)$ to ion motion and inner shell ionization, thereby providing complementary information to a measurement of $A(q)$. The term $D(q)$ provides a new means of comparing damage simulations to experiment, and testing the assumptions that underpin damage models for the single molecule case.

For low diffraction intensities, the dominant error in the calculation of $D(q)$ from experimental data is the error of $\mu_\textrm{pix}(q)$, given by
\begin{align}
\delta \mu_\textrm{pix}(q) = \frac{\sqrt{\mu_\textrm{pix}(q)}}{\sqrt{N_\textrm{DATA}} \sqrt{M(q)}}\;,
\end{align}
where $N_\textrm{DATA}$ is the number of diffraction patterns recorded. The term $M(q)$ is the number of speckles in resolution ring $q$, which is estimated by dividing the circumference of the ring by the expected speckle width $\frac{1}{d}$, where $d$ is the width of the molecule. Assuming $D(q)$ is of the order of one, the error in $D(q)$ goes like $\delta D(q) \approx |\delta \mu_\textrm{pix}(q)|  /  |\mu_\textrm{pix}(q)| $. For the test molecule quoted above and 8 keV photon energy, 2 mJ pulse energy, $100 \times 100$ nm$^2$ spot size at a resolution of $q = 6.67$ nm$^{-1}$, an accuracy of  $\delta D(q) = 0.01$ can be achieved in of the order of $10^3$ patterns, which is an order of magnitude less than the number required to achieve the same resolution in an imaging experiment\cite{Tegze2012}. This analysis could be used to gain early feedback about the data used in an imaging experiment.

\section{Discussion}
\label{sec:discussion}

The results presented on damage noise have implications for the feasibility of determining assembling the 3D diffraction volume from the ensemble of noisy 2D measurements. The data-assembly algorithms use information common to different diffraction measurements to resolve unknown information about molecular orientation. Predicting the level of damage noise in individual 2D diffraction measurements is a first step toward understanding how damage affects these algorithms. The prediction that $SNR_D$ is greater than one even for longer pulse durations ($>$20 fs) is a preliminary indication that damage noise will not prevent data assembly under conditions currently available in experiment. This is because the contribution to the diffraction from the average molecular structure is greater than the shot-to-shot fluctuations of the diffraction, and it is the contribution from the averaged structure that is used to resolve the problem of unknown molecular orientations. That $SNR_{D+N}(q)$ is lower than $SNR_{D}(q)$ by more than an an order of magnitude (see Fig. \ref{fig:pulselength_SNR}) shows that shot noise dominates damage noise. This can be viewed positively because data-assembly algorithms can already cope with very low shot noise levels when assisted by a priori knowledge about the shot noise statistics\cite{LohEMC2009,Fung2009}. However, shot noise applies per pixel and is well understood to be a Poisson process, whereas damage noise applies to features the size of a speckle and the underlying distribution is hard to predict analytically. Detailed studies of the effects of damage on the performance of data assembly algorithms are still required. 

Our study is restricted to spatially uncorrelated damage processes. One significant omission is the expansion of the molecule due to the large electrostatic forces created by the positively charged molecule and the redistribution of trapped electrons. Hydrodynamic simulations have predicted that atoms at the surface can move distances comparable the molecule's size on a time-scale of tens of femtoseconds\cite{HauRiege2004}, while the interior of the molecule moves less in the same time frame, because the trapped electrons redistribute to neutralize the central part of the molecule. The interior atoms will still produce a significant diffraction signal for resolving unknown orientations and assembling the diffraction data. If the surface atoms have moved significantly, they will contribute less to the assembled 3D diffraction data than the interior atoms. If the scattering of surface atoms do prove to reduce relative to the bulk, it is an outstanding question as to how to account for this during structure determination, but modal methods for studying diffraction leave options open \cite{Quiney2011}. 
 
Since damage has been measured in nanocrystallography experiments, it is worth drawing a distinction between damage in crystals and in single molecules. In a crystal, damage ionizes and displaces ions differently in each unit cell, so that the diffraction contains an average over many different damage scenarios. For a single molecule, there is only one damage scenario per measurement and hence we expect a bigger standard deviation of diffraction of single molecules than of nanocrystals. Additionally, nanocrystals are much larger than single molecules, so that the rates at which electrons are trapped is different and the time it takes for a photoelectron to escape is longer. The water that surrounds a nanocrystal injected via a liquid jet \cite{DePonte2008} also contributes to the damage in the form of additional photoelectrons and secondary electrons. It is proposed to inject single molecules via aerosol injection \cite{Bogan2010}, so that they are surrounded by vacuum, because the background water scattering from a liquid jet would dominate the diffraction from the molecule. For these reasons, damage experiments on single molecules, independent of those on crystals, are needed to draw conclusions for single molecule imaging.
 
At the x-ray energies required to reach atomic resolution ($\sim$ 10 keV), Compton scattering becomes another significant source of background scattering\cite{Slowick2014}. The background is predicted to depend on the magnitude of $q$, and would increase the noise level $\sigma_N$ by adding to the right hand side of Eq. \eqref{eq:sigN}. It has been predicted that for for beam intensities currently available at hard x-ray energies, the Compton background only becomes significant at resolutions greater than 2 \AA \cite{Slowick2014}. Hence, Compton scattering is not expected to significantly influence the results presented here.

\section{Conclusion}

We have analyzed shot-to-shot damage-noise fluctuations for single molecule diffraction. For spatially uncorrelated damage processes, there is a clear damage gating effect by which longer pulses measure the same average diffraction contrast as shorter pulses with the same intensity. The results further suggest that pulse energy is more important than pulse duration for maximizing signal to noise for these damage processes. In other words, a pulse 30 fs in duration may be preferable to a sub 5 fs pulse, if the later has an order of magnitude less pulse energy. If both 30 fs and 5 fs pulses have same pulse energy, then the shorter pulse is preferable because damage is reduced, which may be important for damage processes not considered here like the Coulomb explosion. These results provide a preliminary indication that the prospects of resolving molecular orientations to assemble in a 3D diffraction volume in the presence of damage are favorable with data from current facilities. We have also proposed a statistical measure of damage that could be applied experimentally to provide valuable feedback for modeling XFEL damage to single biological molecules.

\appendix
\section{Description of the rate-equations model }
\label{app:model}

We use a damage model based on a rate-equations model\cite{HauRiege2004}, which is extended to include ion diffusion using the methods from a non-local thermal equilibrium plasma model\cite{Scott2001,Caleman2011}. Rates of photoionization are taken from Ref. \cite{Henke1993}, rates of Auger decay were taken from Ref. \cite{McGuire1969} and atomic energy levels were taken from Ref. \cite{Bearden1967}. Secondary impact ionization rates were taken from Refs. \cite{Bell1983,Lennon1988}. Ejected electrons are assumed to be trapped if their kinetic energy exceeds the trapping energy of the ionized molecule\cite{HauRiege2004}. We assume a spherical geometry for this calculation, and this is the only place geometry is included in the calculation. Both photoelectrons and some of the Auger electrons have sufficient energy to escape at early times. All of the trapped electrons are assumed to thermalize on a sub-femtosecond time scale, so that the energy distribution is Maxwell-Boltzmann, but the mean temperature changes with time. We include all ionization states of each element and the electron orbitals for each ionization state were modeled using Slater-type orbitals\cite{Slater1930}.

There are some minor differences between our model and the published models on which it is based. We include all the shells for sulfur (in Ref. \cite{HauRiege2004} it was restricted to 8 electrons). This introduces high energy Auger electrons that are able to escape the molecule under the same conditions as the photoelectrons. We do not consider ionization due to potential lowering, as is done in Ref. \cite{Scott2001}. We also omit the expansion of the molecule under electrostatic forces in order to focus on the spatially uncorrelated motion that is implicated in the self-gating pulse effect. The expansion of a protein molecule has been predicted to affect atoms less than one tenth of the molecule's radius from the surface \cite{HauRiege2004}. These atoms can move several {\AA}ngstr{\"o}m during interaction with the pulse, which will greatly diminish their contribution to the diffraction contrast. The rest of the atoms are only weakly affected by expansion because the trapped electrons effectively neutralize the core, for which we would expect better agreement with the theory presented here.

\section{Derivation of Eq. \eqref{eq:Iav}}
\label{app1}

The intensity of a measurement can be written as
\begin{align}
I(\textbf{q}) = r_e^2 P(\textbf{q}) d\Omega I_0 \Bigg[ &\int \sum_{i=1}^N f_i(q,t)^2 dt + 2 \sum_{i=1}^N \sum_{j=1}^{i-1} \int f_i(q,t) f_j(q,t)
\nonumber \\
&\qquad\qquad \times\cos[ 2 \pi \textbf{q} \cdot (\textbf{R}_i - \textbf{R}_j + \bm{\epsilon}_i(t) - \bm{\epsilon}_j(t) )] dt \Bigg] \;,
\label{eq:app:I}
\end{align}
where the definitions of all terms are given in the main text. We can expand the cosine term as:
\begin{align}
\cos[ 2 \pi \textbf{q} \cdot (\textbf{R}_i - \textbf{R}_j + \bm{\epsilon}_i(t) - \bm{\epsilon}_j(t) )] =& \cos[ 2 \pi \textbf{q} \cdot (\textbf{R}_i - \textbf{R}_j)]\cos[ 2 \pi \textbf{q} \cdot (\bm{\epsilon}_i(t) - \bm{\epsilon}_j(t) )] 
\nonumber \\
&- \sin[ 2 \pi \textbf{q} \cdot (\textbf{R}_i - \textbf{R}_j)]\sin[ 2 \pi \textbf{q} \cdot (\bm{\epsilon}_i(t) - \bm{\epsilon}_j(t) )]
\end{align}
We can further expand the terms that depend upon the displacement as:
\begin{equation}
\cos[ 2 \pi \textbf{q} \cdot (\bm{\epsilon}_i(t) - \bm{\epsilon}_j(t) )]  =  \cos[ 2 \pi \textbf{q} \cdot \bm{\epsilon}_i(t)] \cos[ 2 \pi \textbf{q} \cdot \bm{\epsilon}_j(t)] +  \sin [2 \pi \textbf{q} \cdot \bm{\epsilon}_i(t)] \sin [2 \pi \textbf{q} \cdot \bm{\epsilon}_j(t)] \;.
\end{equation}
The ensemble averages of individual cosine and sine terms over different random displacements are 
\begin{align}
\Big\langle \cos[ 2 \pi \textbf{q} \cdot \bm{\epsilon}_i(t)] \Big\rangle &= \int \cos[ 2 \pi \textbf{q} \cdot \bm{\epsilon}_i(t)] \frac{1}{\sqrt{2\pi} \overline{\bm{\epsilon}}(t)} e^{-\frac{(\bm{q}\cdot\bm{\epsilon}_i(t))^2}{2 \overline{\bm{\epsilon}}(t)^2}} d\bm{\epsilon}_i
\nonumber \\
&= e^{-2\pi^2 q^2 \overline{\bm{\epsilon}}(t)^2}
\label{eq:app:cosav}
\end{align}
and 
\begin{align}
\Big\langle \sin[ 2 \pi \textbf{q} \cdot \bm{\epsilon}_i(t)] \Big\rangle &= \int \sin[ 2 \pi \textbf{q} \cdot \bm{\epsilon}_i(t)] \frac{1}{\sqrt{2\pi} \overline{\bm{\epsilon}}(t)} e^{-\frac{(\bm{q}\cdot\bm{\epsilon}_i(t))^2}{2 \overline{\bm{\epsilon}}(t)^2}} d\bm{\epsilon}_i
\nonumber \\
&= 0 \;.
\label{eq:app:sinav}
\end{align}
We assume that ionization and atomic motion are statistically independent so that
\begin{equation}
\left\langle f_i(q,t) f_j(q,t) \cos[ 2 \pi \textbf{q} \cdot \bm{\epsilon}_i(t)] \right\rangle = \left\langle f_i(q,t) f_j(q,t) \right\rangle \left\langle \cos[ 2 \pi \textbf{q} \cdot \bm{\epsilon}_i(t)] \right\rangle \;.
\end{equation}
We assume that the ionization of different atoms is statistically independent so that 
\begin{equation}
\langle f_i(q,t) f_j(q,t) \rangle = \langle f_i(q,t) \rangle \langle f_j(q,t) \rangle \;,
\end{equation}
if $i \ne j$. We assume that all the atoms of the same element are equivalent statistically, so that averages of $f_i(q,t)$ and $\bm{\epsilon}_i(t)$ are independent of $i$. Combining the above results we get 
\begin{equation}
\Big\langle f_i(q,t) f_j(q,t) \cos[ 2 \pi \textbf{q} \cdot \bm{\epsilon}_i(t)] \cos[ 2 \pi \textbf{q} \cdot \bm{\epsilon}_j(t)] \Big\rangle = \langle f(q,t)\rangle^2 e^{-4\pi^2 q^2 \overline{\bm{\epsilon}}(t)^2} \;.
\label{eq:app:ffcc}
\end{equation}
Substituting Eq. \eqref{eq:app:ffcc} into Eq. \eqref{eq:app:I} leads to Eq. \eqref{eq:Iav}, using the definitions of $A(q)$ and $B(q)$ in Eqs. \eqref{eq:Aav} and \eqref{eq:Bijav} respectively.

\section{Derivation of the variance of $A_i(q)$ : Eq. \eqref{eq:sigA} }
\label{sec:app:sigA}

The standard deviation of the sum of $A_i(q)$ terms in Eq. \ref{eq:I}, denoted by $\sigma_A(q)$, is given by
\begin{align}
\sigma_A^2(q) &= \frac{1}{N} \left[ \left\langle \left[\sum_{i=1}^N A_i(q) \right]^2 \right\rangle - \left\langle \sum_{i=1}^N A_i(q) \right\rangle^2 \right] \;,
\label{eq:sig2A_append}
\end{align}
with 
\begin{equation}
A_i(q) = \int^T_0 f^2_i(q,t) dt \;.
\end{equation}
Equation \eqref{eq:sig2A_append} is scaled the number of atoms to give the contribution per atom. We ignore the $i$ dependence when writing $\sigma_A(q)$ because we assume all atoms of the same element are equivalent. Using the assumption that ionization on different atoms is statistically independent, we can write
\begin{align}
\left\langle \left[\sum_{i=1}^N A_i(q) \right]^2 \right\rangle &= \left\langle \sum_{i=1}^N \int^T_0 f^2_i(q,t) dt \sum_{j=1}^N \int f^2_j(q,t') dt' \right\rangle
\nonumber \\
&=  \sum_{i=1}^N \int^T_0 \langle f^2_i(q,t) f^2_i(q,t') \rangle dt dt' + \sum_{i=1}^N \sum_{j \ne i} \int^T_0 \int^T_0 \langle f^2_i(q,t) \rangle \langle f^2_j(q,t') \rangle dt dt'
\nonumber \\
&= N \int^T_0 \langle f^2_i(q,t) f^2_i(q,t') \rangle dt dt' + N(N-1) \left[ \int^T_0 \langle f^2_i(q,t) \rangle dt \right]^2 \;.
\end{align}
Therefore,
\begin{align}
\sigma_A^2(q) &= \frac{1}{N} \left[ \left\langle \left[\sum_{i=1}^N A_i(q) \right]^2 \right\rangle - \left\langle \sum_{i=1}^N A_i(q) \right\rangle^2 \right]
\nonumber \\
& = \int^T_0 \langle f^2_i(q,t) f^2_i(q,t') \rangle dt dt' - \left[ \int \langle f^2_i(q,t) \rangle dt \right]^2 \;.
\end{align}

\section{Derivation of the variance of $B_{ij}(q)$ : Eq. \eqref{eq:sigB} }
\label{sec:app:sigB}

The term $\sigma_B(q)$ gauges the magnitude of the damage noise fluctuations per atom due to the second term on the right-hand side of Eq. \eqref{eq:I}. Its square is related to the difference between the variance of the second term on the right-hand side of Eq. \eqref{eq:I} and that of the second term on the right-hand side of Eq. \eqref{eq:Iav}, which is given as follows
\begin{align}
\sigma^2_{B}(q) &= \frac{1}{N^2}\left[ \sigma^2_S(q) - \frac{1}{2}(N^2 - N) B^2(q) \right]\;,
\label{eq:sigB_append}
\end{align}
where $\sigma_S(q)$ is defined to be the standard deviation of the second term on r.h.s. of Eq. \eqref{eq:I} and is given by
\begin{equation}
\sigma^2_S(q)  = 4 \sum^N_{i=1} \sum^{i-1}_{j=1} \sum^N_{r=1} \sum^{r-1}_{s=1} \left\langle  B_{ij}(q) B_{rs}(q) \right\rangle \;.
\label{eq:sigS}
\end{equation}
The second term on the right-hand side of Eq. \eqref{eq:app:I} contains terms with the form
\begin{align}
B_{ij}(q) &= \int^T_0 f_i(q,t) f_j(q,t) \cos[ 2 \pi \textbf{q} \cdot (\textbf{R}_i - \textbf{R}_j + \bm{\epsilon}_i(t) - \bm{\epsilon}_j(t) )] dt 
\nonumber \\
 &= B_c(q) \cos[ 2 \pi \textbf{q} \cdot (\textbf{R}_i - \textbf{R}_j)] + B_s(q) \sin[ 2 \pi \textbf{q} \cdot (\textbf{R}_i - \textbf{R}_j)] \;,
\end{align}
where we have defined
\begin{equation}
B_c(q) = \int^T_0 f_i(q,t) f_j(q,t) \cos[ 2 \pi \textbf{q} \cdot (\bm{\epsilon}_i(t) - \bm{\epsilon}_j(t) )]  dt
\end{equation}
and 
\begin{equation}
B_s(q) = \int^T_0 f_i(q,t) f_j(q,t) \sin[ 2 \pi \textbf{q} \cdot (\bm{\epsilon}_i(t) - \bm{\epsilon}_j(t) )]  dt \;.
\end{equation}
Using Eq. \eqref{eq:app:sinav} we can show that 
\begin{equation}
\langle B_s(q) \rangle  = 0 \;,
\end{equation}
and thus write 
\begin{equation}
\langle B(q) \rangle  = \langle B_c(q) \rangle \;.
\end{equation}
We evaluate $\langle B^2_{ij}(q) \rangle$ as a first step to calculating the standard deviation.

\begin{align}
\langle B^2_{ij}(q) \rangle &= \Big\langle \left\{ B_c(q) \cos[ 2 \pi \textbf{q} \cdot (\textbf{R}_i - \textbf{R}_j)] + B_s(q) \sin[ 2 \pi \textbf{q} \cdot (\textbf{R}_i - \textbf{R}_j)] \right\}^2 \Big\rangle
\nonumber \\
&= \langle B^2_c(q) \rangle \langle \cos^2[ 2 \pi \textbf{q} \cdot (\textbf{R}_i - \textbf{R}_j)] \rangle + \langle B^2_s(q) \rangle \langle \sin^2[ 2 \pi \textbf{q} \cdot (\textbf{R}_i - \textbf{R}_j)] \rangle
\nonumber \\
&= \frac{1}{2}\left[ \langle B^2_c(q) \rangle + \langle B^2_s(q) \rangle \right] \;.
\label{eq:app:x2}
\end{align}
Going from the first to the second line of Eq. \eqref{eq:app:x2}, we have used the assumption that the positions of the atoms are random, so that 
\begin{equation}
\Big\langle \cos[ 2 \pi \textbf{q} \cdot (\textbf{R}_i - \textbf{R}_j)] \sin[ 2 \pi \textbf{q} \cdot (\textbf{R}_i - \textbf{R}_j)] \Big\rangle  = 0
\end{equation}
and, in the last line of  Eq. \eqref{eq:app:x2}, we have
\begin{equation}
\langle \cos^2[ 2 \pi \textbf{q} \cdot (\textbf{R}_i - \textbf{R}_j)] \rangle = \langle \sin^2[ 2 \pi \textbf{q} \cdot (\textbf{R}_i - \textbf{R}_j)] \rangle = \frac{1}{2} \;.
\end{equation}
To evaluate Eq. \eqref{eq:app:x2}, we start by evaluating $\langle B^2_c(q) \rangle$ as follows
\begin{align}
\langle B^2_c(q) \rangle = \int^T_0 \int^T_0 &\langle f_i(q,t) f_i(q,t') \rangle \langle f_j(q,t) f_j(q,t') \rangle  
\nonumber \\
& \left\langle\cos[ 2 \pi \textbf{q} \cdot (\bm{\epsilon}_i(t) - \bm{\epsilon}_j(t) )] \cos[ 2 \pi \textbf{q} \cdot (\bm{\epsilon}_i(t') - \bm{\epsilon}_j(t') )] \right\rangle dt dt' \;.
\end{align}
Writing $c_i(t) = \cos[ 2 \pi \textbf{q} \cdot \bm{\epsilon}_i(t)]$, we can write
\begin{align}
\Big\langle \cos[ 2 \pi \textbf{q} \cdot (\bm{\epsilon}_i(t) - \bm{\epsilon}_j(t) )] &\cos[ 2 \pi \textbf{q} \cdot (\bm{\epsilon}_i(t') - \bm{\epsilon}_j(t') )] \Big\rangle 
\nonumber \\
&= \Big\langle \left[ c_i(t) c_j(t) + s_i(t) s_j(t) \right] \left[ c_i(t') c_j(t') + s_i(t') s_j(t') \right]  \Big\rangle
\nonumber \\
&= \langle c_i(t) c_i(t') \rangle \langle c_j(t) c_j(t') \rangle 
\nonumber \\
&\qquad + \langle c_i(t) s_i(t') \rangle \langle c_j(t) s_j(t') \rangle 
\nonumber \\
&\qquad + \langle s_i(t) c_i(t') \rangle \langle s_j(t) c_j(t') \rangle 
\nonumber \\
&\qquad + \langle s_i(t) s_i(t') \rangle \langle s_j(t) s_j(t') \rangle
\nonumber \\
&= \langle c(t) c(t') \rangle^2 + \langle s_i(t) s_i(t') \rangle^2 \;.
\end{align}

The term $\langle c(t) c(t') \rangle$ is given by
\begin{align}
\langle c(t) c(t') \rangle = &\int \int  \cos[ 2 \pi \bm{q} \cdot \bm{\epsilon}(t) ]  \cos[ 2 \pi \bm{q} \cdot \bm{\epsilon}(t') ]  P[\bm{\epsilon}(t),\bm{\epsilon}(t') ] d\bm{\epsilon}(t) d\bm{\epsilon}(t') \;.
\label{eq:integral:o2cos}
\end{align}
The joint probability function is 
\begin{equation}
P[\bm{\epsilon}(t),\bm{\epsilon}(t')] = P[\bm{\epsilon}(t) | \bm{\epsilon}(t')] P[\bm{\epsilon}(t')] \;.
\end{equation}
Assume that $t > t'$. We then assume that the conditional probability is probability of taking a random walk from position $\bm{\epsilon}(t')$ at time $t'$ to position $\bm{\epsilon}(t)$ at time $t$, and takes the form
\begin{equation}
P[\bm{\epsilon}(t) | \bm{\epsilon}(t')] = \frac{1}{(\overline{\bm{\epsilon}}(t,t') \sqrt{2\pi})^3} e^{\frac{- |\bm{\epsilon}_t - \bm{\epsilon}_{t'}|^2 }{2 \overline{\bm{\epsilon}}(t,t')^2}} \;,
\end{equation}
where $\overline{\bm{\epsilon}}(t,t')$ is given by the integral of the diffusion coefficient as a function of time
\begin{equation}
\overline{\bm{\epsilon}}^2(t,t') = 2 N_D \int^t_{t'} d(t'') dt'' \;.
\end{equation}
The term $N_D$ is the number of dimensions, which we will take to be one because we are only interested in diffusion in the direction of the scattering vector. The diffusion coefficient is given by 
\begin{equation}
d(t) = \frac{k_b T(t)}{m \nu(t)} \;,
\end{equation}
where $k_b$ is Boltzmann's constant, $T(t)$ is the ion temperature, $m$ is the ion mass and $\nu(t)$ is the collision frequency.
To evaluate Eq. \eqref{eq:integral:o2cos}, we first write each cosine term as a sum of exponentials
\begin{align}
\cos[ 2 \pi \bm{q} \cdot \bm{\epsilon}(t) ] &= \frac{1}{2} [ e^{2 \pi i \bm{q} \cdot \bm{\epsilon}(t)} + e^{-2 \pi i \bm{q} \cdot \bm{\epsilon}(t)} ]
\nonumber \\
&= \frac{1}{2} \sum^1_{m=0}  e^{ (-1)^m 2 \pi i \bm{q} \cdot \bm{\epsilon}(t)} \;.
\end{align}
We then solve two integrals of the form
\begin{equation}
\int^{\infty}_{-\infty} \sqrt{\frac{a}{\pi}} e^{-a x^2 - bx} dx = e^{\frac{b^2}{4a}} \;.
\end{equation}
The first integral is over $\bm{\epsilon}(t)$, with $a = \frac{1}{2 \overline{\bm{\epsilon}}^2(t,t')}$ and $b = \frac{\bm{\epsilon}(t')}{\overline{\bm{\epsilon}}^2(t,t')} + (-1)^m 2 \pi \bm{q} i$. The argument of the resulting exponent is
\begin{equation}
\frac{b^2}{4a} = \frac{1}{2} \frac{\bm{\epsilon}^2(t')}{\overline{\bm{\epsilon}}^2(t,t')} + (-1)^m 2 \pi  i \bm{q} \cdot \bm{\epsilon}(t') - 2 \pi^2 q^2 \overline{\bm{\epsilon}}^2(t, t') \;.
\end{equation}
The second integral over $\bm{\epsilon} (t')$ has 
\begin{align}
a &= -\frac{1}{2 \overline{\bm{\epsilon}}^2(t,t')} + \frac{1}{2 \overline{\bm{\epsilon}}^2(t,t')} + \frac{1}{2 \overline{\bm{\epsilon}}a^2(t')} = \frac{1}{2 \overline{\bm{\epsilon}}^2(t')}
\nonumber \\
b &= (-1)^m 2 \pi \bm{q} i  + (-1)^n 2 \pi \bm{q} i
\nonumber \\
\frac{b^2}{4a} &= - 2 \pi^2 \overline{\bm{\epsilon}}^2(t') q^2 [(-1)^m + (-1^n)]^2 \;.
\end{align}
The final summation over $m,n = 0,1$ gives the following result for $t > t'$:
\begin{align}
& \int \cos[ 2 \pi \bm{q} \cdot \bm{\epsilon}(t) ] \cos[ 2 \pi \bm{q} \cdot \bm{\epsilon}(t') ]  P[\bm{\epsilon}(t),\bm{\epsilon}(t')] d\bm{\epsilon}(t) d\bm{\epsilon}(t')
\nonumber \\
&=  \frac{1}{2} e^{- 2 \pi^2 q^2 \overline{\bm{\epsilon}}^2(t, t')} [ 1 + e^{- 8 \pi^2 q^2 \overline{\bm{\epsilon}}^2(t')} ] \;.
\label{eq:cos2av}
\end{align}
The corresponding sine integral evaluates to
\begin{align}
& \int \sin[ 2 \pi \bm{q} \cdot \bm{\epsilon}(t) ] \sin[ 2 \pi \bm{q} \cdot \bm{\epsilon}(t') ]  P[\bm{\epsilon}(t),\bm{\epsilon}(t')] d\bm{\epsilon}(t) d\bm{\epsilon}(t') 
\nonumber \\
&=  \frac{1}{2} e^{- 2 \pi^2 q^2 \overline{\bm{\epsilon}}^2(t, t')} [ 1 - e^{- 8 \pi^2 q^2 \overline{\bm{\epsilon}}^2(t')} ] \;.
\label{eq:sin2av}
\end{align}
Adding the cosine and sine integrals, we get
\begin{align}
\langle c(t) c(t') \rangle^2 + \langle s_i(t) s_i(t') \rangle^2 = \frac{1}{2}e^{- 4 \pi^2 q^2 \overline{\bm{\epsilon}}^2(t, t')} [ 1 + e^{- 16 \pi^2 q^2 \overline{\bm{\epsilon}}^2(t')} ] \qquad\qquad (t > t') \;.
\label{eq:ctt_stt}
\end{align}
To complete the evaluation of Eq. \eqref{eq:app:x2}, we still need to evaluate $\langle B^2_s(q) \rangle$ which is given by
\begin{align}
\langle B^2_s(q) \rangle = \int^T_0 \int^T_0 &\langle f_i(q,t) f_i(q,t') \rangle \langle f_j(q,t) f_j(q,t') \rangle
\nonumber \\
& \langle \sin[ 2 \pi \textbf{q} \cdot (\bm{\epsilon}_i(t) - \bm{\epsilon}_j(t) )] \sin[ 2 \pi \textbf{q} \cdot (\bm{\epsilon}_i(t') - \bm{\epsilon}_j(t') )] \rangle dt dt' \;.
\end{align}
This equation can be written in the form
\begin{align}
\langle \sin[ 2 \pi \textbf{q} \cdot (\bm{\epsilon}_i(t) - \bm{\epsilon}_j(t) )] &\sin[ 2 \pi \textbf{q} \cdot (\bm{\epsilon}_i(t') - \bm{\epsilon}_j(t') )] \rangle 
\nonumber \\ 
&= \langle \left[ s_i(t) c_j(t) - c_i(t) s_j(t) \right] \left[ s_i(t') c_j(t') - c_i(t') s_j(t') \right]  \rangle
\nonumber \\
&= \langle s_i(t) s_i(t') \rangle \langle c_j(t) c_j(t') \rangle 
\nonumber \\
&\qquad + \langle c_i(t) c_i(t') \rangle \langle s_j(t) s_j(t') \rangle 
\nonumber \\
&= 2 \langle c_i(t) c_i(t') \rangle \langle s_j(t) s_j(t') \rangle \;.
\end{align}
Using Eqs. \eqref{eq:cos2av} and \eqref{eq:sin2av} we can write this as
\begin{equation}
2 \langle c_i(t) c_i(t') \rangle \langle s_j(t) s_j(t') \rangle = \frac{1}{2} e^{- 4 \pi^2 q^2 \overline{\bm{\epsilon}}^2(t, t')} [ 1 - e^{- 16 \pi^2 q^2 \overline{\bm{\epsilon}}^2(t')} ] \qquad\qquad (t > t') \;.
\end{equation}
We can write the time integrals as
\begin{align}
\langle B^2_c(q) \rangle + \langle B^2_s(q) \rangle = &\int_0^T \int_{t'}^T  \langle f(q,t) f(q,t') \rangle^2 e^{- 4 \pi^2 q^2 \overline{\bm{\epsilon}}^2(t, t')} dt dt' 
\nonumber \\
&+ \int_0^T \int_{0}^{t'}  \langle f(q,t) f(q,t') \rangle^2 e^{- 4 \pi^2 q^2 \overline{\bm{\epsilon}}^2(t', t)} dt dt'
\label{eq:B2c_B2s}
\end{align}
Using the property that $\overline{\bm{\epsilon}}^2(t, t') = -\overline{\bm{\epsilon}}^2(t', t)$, Eq. \eqref{eq:B2c_B2s} can also be written as
\begin{align}
\langle B^2_c(q) \rangle + \langle B^2_s(q) \rangle &= \int_0^T \int_0^T  \langle f(q,t) f(q,t') \rangle^2 e^{- 4 \pi^2 q^2 |\overline{\bm{\epsilon}}^2(t, t')|} dt dt' 
\nonumber \\
&\equiv \langle B^2(q) \rangle \;.
\label{eq:app:BcBscombined}
\end{align}
Using Eqs. \eqref{eq:app:x2} and \eqref{eq:app:BcBscombined} and that $\langle B_{ij} \rangle = 0$, we can calculate the standard deviation of $B_{ij}$ (denoted $\sigma^2_{B_{ij}}(q)$) to be
\begin{align}
\langle B^2_{ij}(q) \rangle &= \frac{1}{2} \int_0^T \int_0^T  \langle f(q,t) f(q,t') \rangle^2 e^{- 4 \pi^2 q^2 |\overline{\bm{\epsilon}}^2(t, t')|} dt dt'  \;.
\end{align}
We have now reached a point where we can evaluate $\sigma_S(q)$, given by Eq. \eqref{eq:sigS}. The averages of terms $\langle B_{ij}(q) B_{rs}(q) \rangle$ are zero unless $i,j=r,s$, because the averages over the positions $\textbf{R}$ equal zero. Therefore,
\begin{align}
\sigma^2_S(q) &= 4 \sum^N_{i=1} \sum^{i-1}_{j=1} \langle B^2_{ij}(q) \rangle
\nonumber \\
&= 4 \frac{N^2 - N}{2} \langle B^2_{ij}(q) \rangle
\nonumber \\
&= (N^2 - N) \int_0^T \int_0^T  \langle f(q,t) f(q,t') \rangle^2 e^{- 4 \pi^2 q^2 |\overline{\bm{\epsilon}}^2(t, t')|} dt dt'
\end{align}
Using this result in Eq. \eqref{eq:sigB_append}, we obtain the following result:
\begin{align}
\sigma^2_{B}(q) &= \frac{1}{N^2}\left[ \sigma^2_S(q) - \frac{1}{2}(N^2 - N) B^2(q) \right]
\nonumber \\
&= \left(1 - \frac{1}{N} \right) \int^T_0 \int^T_0 \Big[ \langle f(q,t) f(q,t') \rangle^2 e^{-4\pi^2 q^2 |\overline{\bm{\epsilon}}^2(t, t')|} 
\nonumber \\
&\qquad\qquad\qquad\qquad\qquad
	- \langle f(q,t) \rangle^2 \langle f(q,t') \rangle^2 e^{-4\pi^2 q^2 \overline{\bm{\epsilon}}(t)^2} e^{-4\pi^2 q^2 \overline{\bm{\epsilon}}(t')^2} \Big] dt dt' \;.
\end{align}
Assuming that N is large, the term of $\frac{1}{N}$ can be ignored.

\section*{Acknowledgements}

HMQ and AVM acknowledge funding from the Australian Research Council via its Centres of Excellence and Discovery Early Career Researcher Award (DE140100624) programmes. We are grateful to Jochen K{\"u}pper for helpful feedback.

\bibliographystyle{unsrt}
\bibliography{damage_effects}

\begin{thebibliography}{10}

\bibitem{Neutze2000}
R.~Neutze, W.~Wouts, D.~{van der Spoel}, E.~Weckert, and J.~Hajdu.
\newblock Potential for biomolecular imaging with femtosecond x-ray pulses.
\newblock {\em Nature}, 406:752--757, 2000.

\bibitem{LCLS}
P.~Emma, R.~Akre, J.~Arthur, R.~M. Bionta, C.~Bostedt, J.~Bozek, A.~Brachmann,
  P.~H. Bucksbaum, R.~Coffee, F.~J. Decker, Y.~Ding, D.~Dowell, S.~Edstrom,
  A.~Fisher, J.~Frisch, S.~Gilevich, J.~Hastings, G.~Hays, P.~Hering, Z.~Huang,
  R.~Iverson, H.~Loos, M.~Messerschmidt, A.~Miahnahri, S.~Moeller, H.-D. Nuhn,
  D.~Pile, D.~Ratner, J.~Rzepiela, D.~Schultz, T.~Smith, P.~Stefan,
  H.~Tompkins, J.~Turner, J.~Welch, W.~White, J.~Wu, G.~Yocky, , and J.~N.
  Galayda.
\newblock First lasing and operation of an angstrom-wavelength free-electron
  laser.
\newblock {\em Nat. Photonics}, 4:641--648, 2010.

\bibitem{Ishikawa2012}
T.~Ishikawa, H.~Aoyagi, T.~Asaka, Y.~Asano, N.~Azumi, T.~Bizen, H.~Ego,
  K.~Fukami, T.~Fukui, Y.~Furukawa, S.~Goto, H.~Hanaki, T.~Hara, T.~Hasegawa,
  T.~Hatsui, A.~Higashiya, T.~Hirono, N.~Hosoda, M.~Ishii, T.~Inagaki,
  Y.~Inubushi, T.~Itoga, Y.~Joti, M.~Kago, T.~Kameshima, H.~Kimura,
  Y.~Kirihara, A.~Kiyomichi, T.~Kobayashi, C.~Kondo, T.~Kudo, H.~Maesaka, X.~M.
  Maréchal, T.~Masuda, S.~Matsubara, T.~Matsumoto, T.~Matsushita, S.~Matsui,
  M.~Nagasono, N.~Nariyama, H.~Ohashi, T.~Ohata, T.~Ohshima, S.~Ono, Y.~Otake,
  C.~Saji, T.~Sakurai, T.~Sato, K.~Sawada, T.~Seike, K.~Shirasawa, T.~Sugimoto,
  S.~Suzuki, S.~Takahashi, H.~Takebe, K.~Takeshita, K.~Tamasaku, H.~Tanaka,
  R.~Tanaka, T.~Tanaka, T.~Togashi, K.~Togawa, A.~Tokuhisa, H.~Tomizawa,
  K.~Tono, S.~Wu, Makina yabashi, M.~Yamaga, A.~Yamashita, K.~Yanagida,
  C.~Zhang, T.~Shintake, H.~Kitamura, , and N.~Kumagai.
\newblock A compact x-ray free-electron laser emitting in the
  sub-{\r{a}}ngstr{\"{o}}m region.
\newblock {\em Nat. Photonics}, 6:540--544, 2012.

\bibitem{Caleman2009}
C.~Caleman, C.~Ortiz, E.~Marklund, F.~Bultmark, M.~Gabrysch, F.~G. Parak,
  J.~Hajdu, and M.~Klintenbergand~N. Timneanu.
\newblock Radiation damage in biological material: Electronic properties and
  electron impact ionization in urea.
\newblock {\em Europhys. Lett.}, 85:18005, 2009.

\bibitem{Caleman2011}
Carl Caleman, Magnus Bergh, Howard~A. Scott, John~C.H. Spence, Henry~N.
  Chapman, and Nicusor Timneanu.
\newblock Simulations of radiation damage in biomolecular nanocrystals induced
  by femtosecond x-ray pulses.
\newblock {\em J. Mod. Optic.}, 58:1486--1497, 2011.

\bibitem{Chapman2011}
Henry~N. Chapman, Petra Fromme, Anton Barty, Thomas~A. White, Richard~A.
  Kirian, Andrew Aquila, Mark~S. Hunter, Joachim Schulz, Daniel~P. DePonte, Uwe
  Weierstall, R.~Bruce Doak, Filipe R. N.~C. Maia, Andrew~V. Martin, Ilme
  Schlichting, Lukas Lomb, Nicola Coppola, Robert~L. Shoeman, Sascha~W. Epp,
  Robert Hartmann, Daniel Rolles, Artem Rudenko, Lutz Foucar, Nils Kimmel,
  Georg Weidenspointner, Peter Holl, Mengning Liang, Miriam Barthelmess, Carl
  Caleman, S$\acute{e}$bastien Boutet, Michael~J. Bogan, Jacek Krzywinski,
  Christoph Bostedt, Sa$\check{s}$a Bajt, Lars Gumprecht, Benedikt Rudek,
  Benjamin Erk, Carlo Schmidt, Andr$\acute{e}$ H{\"o}mke, Christian Reich,
  Daniel Pietschner, Lothar Str{\"u}der, G{\"u}nter Hauser, Hubert Gorke,
  Joachim Ullrich, Sven Herrmann, Gerhard Schaller, Florian Schopper, Heike
  Soltau, Kai-Uwe K{\"u}hnel, Marc Messerschmidt, John~D. Bozek, Stefan~P.
  Hau-Riege, Matthias Frank, Christina~Y. Hampton, Raymond~G. Sierra, Dmitri
  Starodub, Garth J.Williams, Janos Hajdu, Nicusor Timneanu, M.~Marvin Seibert,
  Jakob Andreasson, Andrea Rocker, Olof J{\"o}nsson, Martin Svenda, Stephan
  Stern, Karol Nass, Robert Andritschke, Claus-Dieter~Schr{\"o} ter, Faton
  Krasniqi, Mario Bott, Kevin~E. Schmidt, XiaoyuWang, Ingo Grotjohann, James~M.
  Holton, Thomas R.~M. Barends, Richard Neutze, Stefano Marchesini, Raimund
  Fromme, Sebastian Schorb, Daniela Rupp, Marcus Adolph, Tais Gorkhover, Inger
  Andersson, Helmut Hirsemann, Guillaume Potdevin, Heinz Graafsma, Bj{\"o}rn
  Nilsson, and John C.~H. Spence.
\newblock Femtosecond x-ray protein nanocrystallography.
\newblock {\em Nature}, 470:73--77, 2011.

\bibitem{Boutet2012}
S.~Boutet, L.~Lomb, G.~J. Williams, T.~R.~M. Barends, A.~Aquila, R.~B. Doak,
  U.~Weierstall, D.~P. DePonte, J.~F. Steinbrener, R.~L. Shoeman,
  M.~Messerschmidt, A.~Barty, T.~A. White, S.~Kassemeyer, R.~A. Kirian, M.~M.
  Seibert, P.~A. Montanez, C.~Kenney, R.~Herbst, P.~Hart, J.~Pines, G.~Haller,
  S.~M. Gruner, H.~T. Philipp, M.~W. Tate, M.~Hromalik, L.~J. Koerner, N.~van
  Bakel, J.~Morse, W.~Ghonsalves, D.~Arnlund, M.~J. Bogan, C.~Caleman,
  R.~Fromme, C.~Y. Hampton, M.~S. Hunter, L.~C. Johansson, G.~Katona,
  C.~Kupitz, M.~Liang, A.~V. Martin, K.~Nass, L.~Redecke, F.~Stellato,
  N.~Timneanu, D.~Wang, N.~A. Zatsepin, D.~Schafer, J.~Defever, R.~Neutze,
  P.~Fromme, J.~C.~H. Spence, H.~N. Chapman, and I.~Schlichting.
\newblock High-resolution protein structure determination by serial femtosecond
  crystallography.
\newblock {\em Science}, 337:362--364, 2012.

\bibitem{Barty2012}
Anton Barty, Carl Caleman, Andrew Aquila, Nicusor Timneanu, Lukas Lomb,
  Thomas~A. White, Jakob Andreasson, David Arnlund, Sasa Bajt, Thomas R.~M.
  Barends, Miriam Barthelmess, Michael~J. Bogan, Christoph Bostedt, John~D.
  Bozek, Ryan Coffee, Nicola Coppola, Jan Davidsson, Daniel~P. DePonte,
  R.~Bruce Doak, Tomas Ekeberg, Veit Elser, Sascha~W. Epp, Benjamin Erk, Holger
  Fleckenstein, Lutz Foucar, Petra Fromme, Heinz Graafsma, Lars Gumprecht,
  Janos Hajdu, Christina~Y. Hampton, Robert Hartmann, Andreas Hartmann, Günter
  Hauser, Helmut Hirsemann, Peter Holl, Mark~S. Hunter, Linda Johansson,
  Stephan Kassemeyer, Nils Kimmel, Richard~A. Kirian, Mengning Liang, Filipe R.
  N.~C. Maia, Erik Malmerberg, Stefano Marchesini, Andrew~V. Martin, Karol
  Nass, Richard Neutze, Christian Reich, Daniel Rolles, Benedikt Rudek, Artem
  Rudenko, Howard Scott, Ilme Schlichting, Joachim Schulz, M.~Marvin Seibert,
  Robert~L. Shoeman, Raymond~G. Sierra, Heike Soltau, John C.~H. Spence,
  Francesco Stellato, Stephan Stern, Lothar Str{\:u}der, Joachim Ullrich,
  X.~Wang, Georg Weidenspointner, Uwe Weierstall, Cornelia~B. Wunderer, and
  Henry~N. Chapman.
\newblock Self-terminating diffraction gates femtosecond x-ray
  nanocrystallography measurements.
\newblock {\em Nature Photonics}, 6:35--40, 2012.

\bibitem{HauRiege2005}
Stefan Hau-Riege, Richard London, Gosta Huldt, and Henry Chapman.
\newblock Pulse requirements for x-ray diffraction imaging of single biological
  molecules.
\newblock {\em Phys. Rev. E.}, 71:061919, 2005.

\bibitem{Ziaja2012}
B~Ziaja, H~N Chapman, R~F{\"a}ustlin, S~Hau-Riege, Z~Jurek, A~V Martin,
  S~Toleikis, F~Wang, E~Weckert, and R~Santra.
\newblock Limitations of coherent diffractive imaging of single objects due to
  their damage by intense x-ray radiation.
\newblock {\em New Journal of Physics}, 14:115015, 2012.

\bibitem{Quiney2011}
Harry~M. Quiney and Keith~A. Nugent.
\newblock Biomolecular imaging and electronic damage using x-ray free-electron
  lasers.
\newblock {\em Nat. Phys.}, 7:142--146, 2011.

\bibitem{LohEMC2009}
Ne-Te~Duane Loh and Veit Elser.
\newblock Reconstruction algorithm for single-particle diffraction imaging
  experiments.
\newblock {\em Phys. Rev. E}, 80:026705, 2009.

\bibitem{Fung2009}
Russell Fung, Valentin Shneerson, Dilano~K. Saldin, and Abbas Ourmazd.
\newblock Structure from fleeting illumination of faint spinning objects in
  flight.
\newblock {\em Nat. Phys.}, 5:64--67, 2009.

\bibitem{Giannakis2012}
D.~Giannakis, P.~Schwander, and A.~Ourmazd.
\newblock The symmetries of image formation by scattering. i. theoretical
  framework.
\newblock {\em Opt. Express}, 20:12799--12826, 2012.

\bibitem{Kassemeyer2013}
Stephan Kassemeyer, Aliakbar Jafarpour, Lukas Lomb, Jan Steinbrener, Andrew~V.
  Martin, and Ilme Schlichting.
\newblock Optimal mapping of x-ray laser diffraction patterns into three
  dimensions using routing algorithms.
\newblock {\em Phys. Rev. E}, 88:042710, 2013.

\bibitem{Tegze2012}
M~Tegze and G~Bortel.
\newblock {Atomic structure of a single large biomolecule from diffraction
  patterns of random orientations}.
\newblock {\em Journal of Structural Biology}, 179:41--45, 2012.

\bibitem{Marchesini2007}
S.~Marchesini.
\newblock A unified evaluation of iterative projection algorithms for phase
  retrieval.
\newblock {\em Rev. Sci. Instrum.}, 78(1):011301, 2007.

\bibitem{Maia2009}
Filipe R. N.~C. Maia, Tomas Ekeberg, Nicusor Tîmneanu, David van~der Spoel, and
  Janos Hajdu.
\newblock Structural variability and the incoherent addition of scattered
  intensities in single-particle diffraction.
\newblock {\em Phys. Rev. E}, 80:031905, 2009.

\bibitem{Kupper2014}
Jochen K\"upper, Stephan Stern, Lotte Holmegaard, Frank Filsinger, Arnaud
  Rouz\'ee, Artem Rudenko, Per Johnsson, Andrew~V. Martin, Marcus Adolph,
  Andrew Aquila, Sasa Bajt, Anton Barty, Christoph Bostedt, John Bozek, Carl
  Caleman, Ryan Coffee, Nicola Coppola, Tjark Delmas, Sascha Epp, Benjamin Erk,
  Lutz Foucar, Tais Gorkhover, Lars Gumprecht, Andreas Hartmann, Robert
  Hartmann, G\"unter Hauser, Peter Holl, Andre H\"omke, Nils Kimmel, Faton
  Krasniqi, Kai-Uwe K\"uhnel, Jochen Maurer, Marc Messerschmidt, Robert
  Moshammer, Christian Reich, Benedikt Rudek, Robin Santra, Ilme Schlichting,
  Carlo Schmidt, Sebastian Schorb, Joachim Schulz, Heike Soltau, H.~Spence,
  John~C.\, Dmitri Starodub, Lothar Str\"uder, Jan Th\o{}gersen, J.~Vrakking,
  Marc~J.\, Georg Weidenspointner, Thomas~A. White, Cornelia Wunderer, Gerard
  Meijer, Joachim Ullrich, Henrik Stapelfeldt, Daniel Rolles, and Henry~N.
  Chapman.
\newblock X-ray diffraction from isolated and strongly aligned gas-phase
  molecules with a free-electron laser.
\newblock {\em Phys. Rev. Lett.}, 112:083002, Feb 2014.

\bibitem{Stern2014}
S.~Stern, L.~Holmegaard, F.~Filsinger, A.~Rouzee, A.~Rudenko, P.~Johnsson,
  A.~V. Martin, A.~Barty, C.~Bostedt, J.~Bozek, R.~Coffee, S.~Epp, B.~Erk,
  L.~Foucar, R.~Hartmann, N.~Kimmel, K-U. Kuhnel, J.~Maurer, M.~Messerschmidt,
  B.~Rudek, D.~Starodub, J.~Thogersen, G.~Weidenspointner, T.~A. White,
  H.~Stapelfeldt, D.~Rolles, H.~N. Chapman, and J.~Kupper.
\newblock Toward atomic resolution diffractive imaging of isolated molecules
  with x-ray free-electron lasers.
\newblock {\em Faraday Discuss.}, 171:393--418, 2014.

\bibitem{Lorenz2012}
U.~Lorenz, N.~M. Kabachnik, E.~Weckert, and I.~A. Vartanyants.
\newblock Impact of ultrafast electronic damage in single-particle x-ray
  imaging experiments.
\newblock {\em Phys. Rev. E}, 86:051911, 2012.

\bibitem{HauRiege2004}
S.~P. Hau-Riege, R.~A. London, and A.~Sz{\~{o}}ke.
\newblock Dynamics of biological molecules irradiated by short x-ray pulses.
\newblock {\em Phys. Rev. E}, 69:051906, 2004.

\bibitem{Jurek2004b}
Z.~Jurek, G.~Oszl{$\acute{a}$}nyi, and G.~Faigel.
\newblock Imaging atom clusters by hard x-ray free-electron lasers.
\newblock {\em Europhys. Lett.}, 65:491--497, 2004.

\bibitem{Slater1930}
J.C. Slater.
\newblock Atomic shielding constants.
\newblock {\em Phys. Rev.}, 36:57--64, 1930.

\bibitem{Huldt2003}
G.~Huldt, A.~Szoke, and J.~Hajdu.
\newblock Diffraction imaging of single particles and biomolecules.
\newblock {\em J. Struct. Biol.}, 144:219, 2003.

\bibitem{Scott2001}
Howard~A. Scott.
\newblock Cretin - a radiative transfer capability for laboratory plasmas.
\newblock {\em Journal of Quantitative Spectroscopy \& Radiative Transfer},
  71:689--701, 2001.

\bibitem{Rostrom1999}
A.A. Rostrom and C.V. Robinson.
\newblock Detection of the intact groel chaperonin assembly by mass
  spectrometry.
\newblock {\em J. Am. Chem. Soc.}, 121:4718--4719, 1999.

\bibitem{Kirian2012}
R.~A. Kirian.
\newblock Structure determination through correlated fluctuations in x-ray
  scattering.
\newblock {\em J. Phys. B: At. Mol. Opt. Phys.}, 45:223001, 2012.

\bibitem{DePonte2008}
D.~P. DePonte.
\newblock {\em Journal of Physics D: Applied Physics}, 41:195505, 2008.

\bibitem{Bogan2010}
M.~J. Bogan, D.~Starodub, C.~Hampton, and R.~G. Sierra.
\newblock Single-particle coherent diffractive imaging with a soft x-ray free
  electron laser: towards soot aerosol morphology.
\newblock {\em J. Phys. B-At. Mol. Op.}, 43:194013, 2010.

\bibitem{Slowick2014}
J~M Slowik, S-K Son, G~Dixit, Z~Jurek, and R~Santra.
\newblock Incoherent x-ray scattering in single molecule imaging.
\newblock {\em New Journal of Physics}, 16:073042, 2014.

\bibitem{Henke1993}
B.~L. Henke, E.~M. Gullikson, and J.~C. Davis.
\newblock {\em At. Data Nucl. Data Tables}, 54:181--342, 1993.

\bibitem{McGuire1969}
E.~J. McGuire.
\newblock K-shell auger transition rates and fluorescence yields for elements
  be-ar.
\newblock {\em Phys. Rev.}, 185:20--30, 1969.

\bibitem{Bearden1967}
J.A. Bearden and J.F. Burr.
\newblock Reevaluation of x-ray atomic energy levels.
\newblock {\em Rev Mod Phys}, 39:125--142, 1967.

\bibitem{Bell1983}
K.~L. Bell, H.~B. Gilbody, J.~G. Hughes, A.~E. Kingston, and F.~J. Smith.
\newblock Recommended data on the electron impact ionization of light atoms and
  ions.
\newblock {\em J. Phys. Chem. Ref. Data}, 12:891--916, 1983.

\bibitem{Lennon1988}
M.~A. Lennon, K.~L. Bell, H.~B. Gilbody, J.~G. Hughes, A.~E. Kingston, M.~J.
  Murray, and F.~J. Smith.
\newblock {\em J. Phys. Chem. Ref. Data}, 17:1285--1363, 1988.

\end{thebibliography}


\end{document}